\documentclass[acmsmall]{acmart}

\usepackage{listings}
\usepackage{hyperref}
\usepackage{graphicx}
\usepackage{csvsimple}
\usepackage{subcaption}
\usepackage{comment}
\usepackage{xcolor}
\usepackage{color, colortbl}
\usepackage{url}
\usepackage{multirow}
\usepackage{amsmath}
\usepackage{rotating}
\usepackage[T1]{fontenc}
\usepackage{soul}
\usepackage{tikz}
\usepackage{makecell}

\definecolor{Gray}{gray}{0.8}
\acmDOI{}
\renewcommand\footnotetextcopyrightpermission[1]{} 

\begin{document}

\title{Metadata Analysis of Open Educational Resources}\thanks{Preprint version -- This paper has been accepted to be published in the 11th International Learning Analytics and Knowledge (LAK'2021), April 12--16, 2021. ACM}  

\author{Mohammadreza Tavakoli}
\affiliation{
\institution{German National Library of Science and Technology (TIB)}
\country{Germany}}
\email{reza.tavakoli@tib.eu}

\author{Mirette Elias}
\email{melias@uni-bonn.de}
\affiliation{
\institution{Fraunhofer IAIS and University of Bonn}
\country{Germany}
}

\author{G\'abor Kismihok}
\affiliation{%
\institution{German National Library of Science and Technology (TIB)}
\country{Germany}}
\email{gabor.kismihok@tib.eu}

\author{S\"oren Auer}
\affiliation{%
\institution{German National Library of Science and Technology (TIB)}
\country{Germany}}
\email{soeren.auer@tib.eu}

\renewcommand{\shortauthors}{Tavakoli and Elias, et al.}

\begin{abstract}
Open Educational Resources (OERs) are openly licensed educational materials that are widely used for learning. Nowadays, many online learning repositories provide millions of OERs. Therefore, it is exceedingly difficult for learners to find the most appropriate OER among these resources. Subsequently, the precise OER metadata is critical for providing high-quality services such as search and recommendation. Moreover, metadata facilitates the process of automatic OER quality control as the continuously increasing number of OERs makes manual quality control extremely difficult. This work uses the metadata of \textit{8,887} OERs to perform an exploratory data analysis on OER metadata. Accordingly, this work proposes metadata-based scoring and prediction models to anticipate the quality of OERs. Based on the results, our analysis demonstrated that OER metadata and OER content qualities are closely related, as we could detect high-quality OERs with an accuracy of \textbf{94.6}\%. Our model was also evaluated on \textit{884} educational videos from Youtube to show its applicability on other educational repositories.
\end{abstract}




\maketitle

\section{Introduction}
Open Educational Resources (OERs) play a key role in informal education these days. There are many OER repositories (e.g., MIT\footnote{\url{https://ocw.mit.edu/}}, edX\footnote{\url{https://www.edx.org/}}, Khan Academy\footnote{\url{https://www.khanacademy.org/}}) hosting and launching millions of OERs under Creative Common license\footnote{\url{https://creativecommons.org/}} on a daily basis. However, the lack of high-quality services, such as OER search and recommendation systems, limit the use of OERs~\cite{chicaiza2017recommendation, tavakoli2020labour, tavakoli2020oer}. In order to provide such services, high-quality metadata that describe OERs thoroughly and reliably are essential~\cite{tavakoli2020quality}.
Although most of the OER repositories are using standardized metadata definitions (e.g., IEEE Standard for Learning Object Metadata (LOM)~\cite{ieee20021484} and Learning Resource Metadata Initiative (LRMI)~\cite{lrmi}) to improve open educational services, the lack or low-quality of metadata still limits the performance of these initiatives~\cite{kiraly2018measuring,ochoa2009automatic}.
Furthermore, OERs vary in terms of a large number of important features (from a learner point of view), like levels of education, topics or vocation. OERs also come in large numbers of different formats and languages. Therefore, it has become inevitable these days to put more emphasis on assessing and controlling the quality of OERs, in which OER metadata should play a prominent role. If OER metadata is created as generic part of the OER quality control processes, automatic metadata analysis may significantly improve the evaluation of OERs. This is not the case currently, as very often only manual methods are used to validate both the quality of OER content and metadata~\cite{Tani2013}, which are time consuming and unscalable solutions~\cite{ochoa2009automatic}.
Although, there are some attempts to automatize quality assessment of metadata \cite{ochoa2009automatic,Trippel2014}, these only focus on the criteria definitions and metrics to evaluate already existing OER metadata~\cite{bruce2004continuum,Ochoa2006,romero2019proposal} without building an intelligent model or models to predict the quality of OERs based on metadata.
Based on our assumption that the quality of OER metadata has tight relationship with the quality of OER content, in this paper we discuss the details of our exploratory data analysis on the metadata of \emph{8,887} OERs from \textit{SkillsCommons}\footnote{\url{http://skillscommons.org}}. This was done in order to provide insights about: (1) the quality of metadata in existing OERs; (2) the effect of quality control on metadata quality; (3) building metadata-based scoring and prediction models to anticipate the quality of OERs; and finally, (4) we evaluated our proposed models by using the metadata of \emph{884} OERs from \textit{YouTube}\footnote{\url{https://www.youtube.com/}}, to demonstrate the general nature of our proposed approach, by applying it to different types of educational resources and repositories. 

The article is organized as follows: Section \ref{sec-related} discusses the state-of-the-art of assessing the quality of OER metadata and also OER content using metadata. Section \ref{sec-method} explains our steps when it comes to data collection, analysis, and the proposed approach of metadata scoring and prediction of OER quality. Section \ref{sec-validation} shares the results of applying our model on Youtube educational videos in order to validate our proposed approach. Finally, Section \ref{sec-discussion} discusses our results and Section \ref{sec-conclusion} drives the conclusion and showcases our future work on this topic.

\section{Related Work}\label{sec-related}
OER metadata is important not only to aid learners in finding relevant content among large amount of OERs, but also to indicate OER quality~\cite{elias2020ectel}. Most of the literature about OER metadata quality focused on metadata records and their data values~\cite{phillips2020using}. In this section, we review the related body of OER metadata literature, in terms of: 1) research defining dimensions and metrics for metadata, and 2) approaches that improve the quality of metadata.

\subsection{Defining Dimensions and Metrics for Metadata}
Currently, the quality of OER metadata has been determined in terms of the following dimensions:~\textit{completeness, accuracy, provenance, conformance to expectations, logical consistency and coherence, timeliness}, and \textit{accessibility}~\cite{bruce2004continuum}. Ochoa and Duval~\cite{Ochoa2006} have converted those dimensions into a set of calculated metrics, which have been reused by most of the researchers addressing quality of OER metadata~\cite{elias2020ectel}. They also partially evaluated their metrics (i.e., completeness, accuracy) on a list of \textit{425} OERs from the ARIADNE Learning Object Repository~\cite{ochoa2009automatic}.
Moreover, \cite{pelaez2017metadata} has evaluated the completeness and consistency of OERs metadata by building their calculation on Ochoa and Duval’s metrics~\cite{Ochoa2006}. They evaluated consistency of metadata elements values with respect to the standardized domain values (e.g., Language should be according to \textit{ISO 639-111} language standard).

\subsection{Improving the Quality of Metadata} 
To have high-quality metadata, some methods have been developed in order to help authors and experts in providing metadata for OERs. A process for improving the metadata quality of OERs was developed to support domain experts with metadata creation; the process introduces qualitative methods (e.g., online peer review of metadata) and tools (e.g., metadata quality assessment grid) in the various phases when it comes to populating metadata in OER repositories \cite{palavitsinis2014metadata}.
Moreover, a higher level of metadata quality analysis was applied to help metadata creators to assess and improve the quality of metadata \cite{phillips2020using}. They exploit linked open data to discover and analyze connectivity between metadata records. Accordingly, they used network statistics (e.g., density of graph) to calculate the relationship between the metadata records in terms of their attributes (e.g. subject) and values. Their study was applied on six large digital library collections and they discussed several improvements that can help users find related resources.

\subsection{Lessons Learned}
Based on the state-of-the-art, although there are several attempts regarding assessing and improving OER metadata, most of these efforts are either conceptual \cite{romero2019proposal}, or focusing only on a few dimensions \cite{margaritopoulos2012quantifying,romero2018exploring}. Furthermore, currently there is no scalable solutions~\cite{ochoa2009automatic} available, which limits the capability of existing approaches, when it comes to OER metadata quality assessment and improvement~\cite{gavrilis2015measuring}. 
Therefore, it is clear that there is a significant need for automatic and intelligent metadata quality assessment in order to improve the discoverability, usability, and reusability of OERs \cite{gavrilis2015measuring}. Subsequently, a recently brief preliminary analysis was conducted on the current state of OER metadata in order to establish a quality prediction model \cite{tavakoli2020quality}. As a result, we conclude that:~(1)~it is worthwhile and timely to consider analyzing OER metadata to improve OER-based services; and (2) there is a lack of intelligent prediction model that evaluates the quality of OERs based on their metadata to facilitate the quality control. 
For the above mentioned reasons, in this paper, we attempt to follow-up, extend and evaluate the OER metadata quality prediction model suggested by \cite{tavakoli2020quality}, by using a video content based OER dataset, consisting of educational videos from \textit{Youtube}. This was done in order to show the scalability and the generalizability of the proposed approach. Accordingly, the main objectives of this paper are:
\begin{enumerate}
    \item Conducting an exploratory data analysis on large amount of OER metadata.
    \item Building a scoring model in a data-driven approach that helps OER repositories and OER authors to evaluate and improve the quality of their OER metadata.
    \item Predicting the quality of OERs based on their metadata, which should guide automatic quality control processes and ultimately result in higher OER quality.
    \item Evaluating our quality prediction model by applying it on educational videos retrieved from \textit{Youtube}.
\end{enumerate}

\section{Data Collection and Research Method}\label{sec-method}

\subsection{Data Collection}
We have used two datasets to analyze the OERs metadata and evaluate our model. The \textit{SkillsCommons} dataset was used to analyze and train our machine learning model and the \textit{YouTube} dataset was used to evaluate our prediction model. 
\subsubsection{SkillsCommons}
For analyzing the OERs metadata and building the quality prediction model, we built an OER metadata dataset after retrieving all search results for the terms \textit{Information Technology} and \textit{Health Care} via the SkillsCommons platform API \cite{tavakoli2020quality}. The dataset contains \textit{8,887} OERs metadata\footnote{Our \emph{SkillsCommons} dataset is available on: \url{https://github.com/rezatavakoli/ICALT2020_metadata}}. The OER metadata in our sample included the following fields: url, title, description, educational type, date of availability, date of issuing, subject list, target audience-level, time required to finish, accessibilities, language list, and quality control (i.e., a categorical value that shows if a particular OER went through a quality control or not). It should be mentioned that the \emph{quality control} field means manual quality control, and it has been set to \textbf{with control} if an OER had at least one inspection regarding the Quality of Subject Matter, and at least one inspection regarding the Quality of Online/Hybrid Course Design, otherwise it is set to \textbf{without control}.

\subsubsection{Youtube} To evaluate our proposed model, we selected 16 topics, which are defined by \cite{tavakoli2020recommender} as \textit{Information Technology} related search keywords. Moreover, we randomly selected another 16 topics from \cite{hot_topics} as \textit{Health Care} related search terms. Afterwards, for each of the 32 selected topics in the areas of \textit{Information Technology} and \textit{Health Care}, top videos in Youtube search results were collected\footnote{Our \emph{Youtube} dataset is available on: \url{https://github.com/rezatavakoli/LAK21_metadata}} using \textit{Pafy} python library\footnote{\url{https://pypi.org/project/pafy/}}. In a Youtube search, the number of top videos appearing in search results depends on the search query topic, and therefore, we can be confronted by different number of videos as top results. However, we collected at least 10 videos per each search term. At the end, \textit{884} Youtube educational videos were collected for our evaluation step\footnote{For the current version, we used openly available videos, but we disregarded the type of license for our analysis. Nevertheless, licensing will obviously play a role in future implementations.}. The video metadata includes the following fields: url, title, description, number of dislikes, length, number of likes, rating, subject list, and number of views.

\subsection{Exploratory Analysis of OER Metadata}
\begin{figure*}[h]
\centering
  \begin{subfigure}{0.4\textwidth}
    \includegraphics[width=6cm]{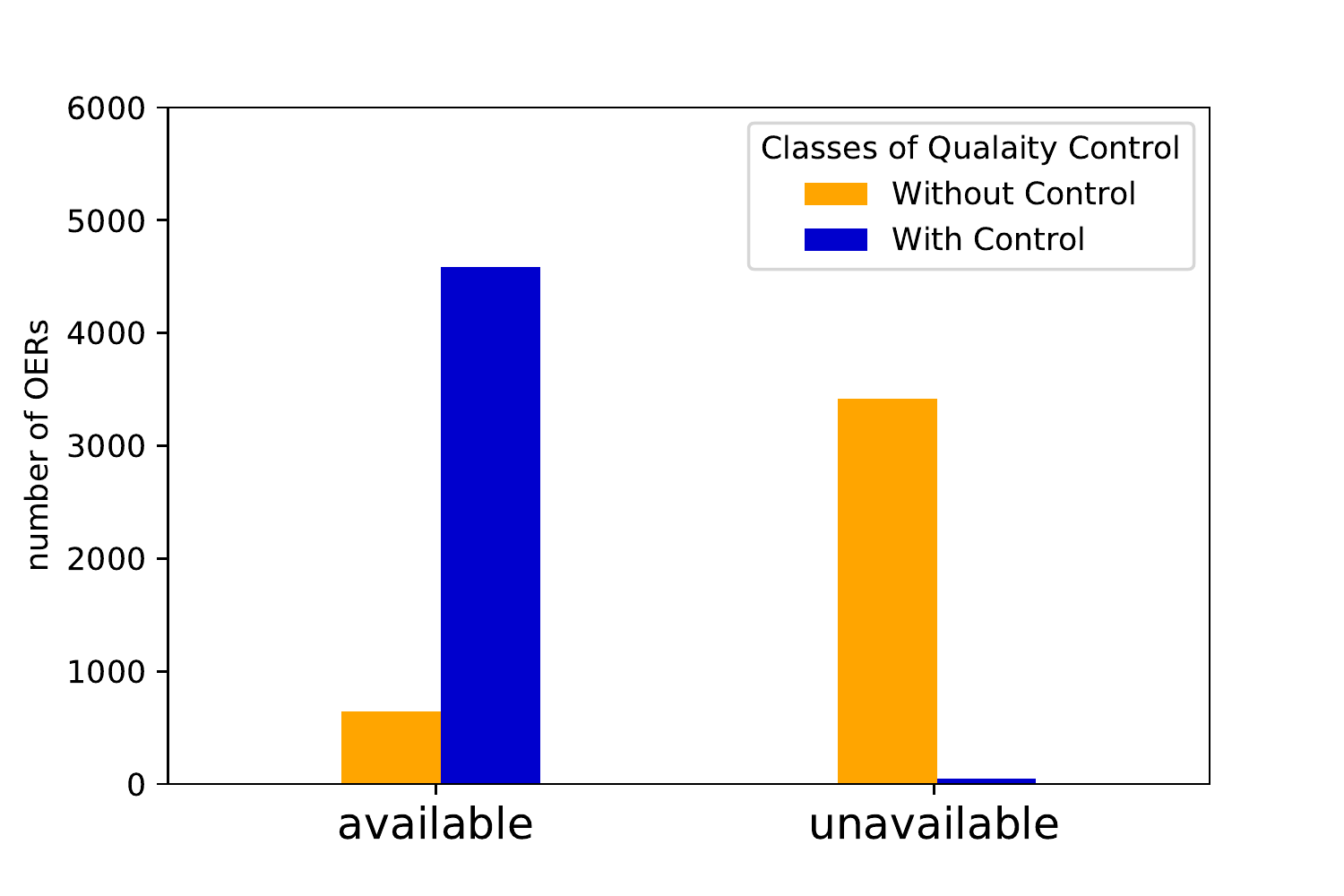}
    \caption{\textit{Level} metadata}
    \label{fig-level}
  \end{subfigure}
  \begin{subfigure}{0.4\textwidth}
    \includegraphics[width=6cm]{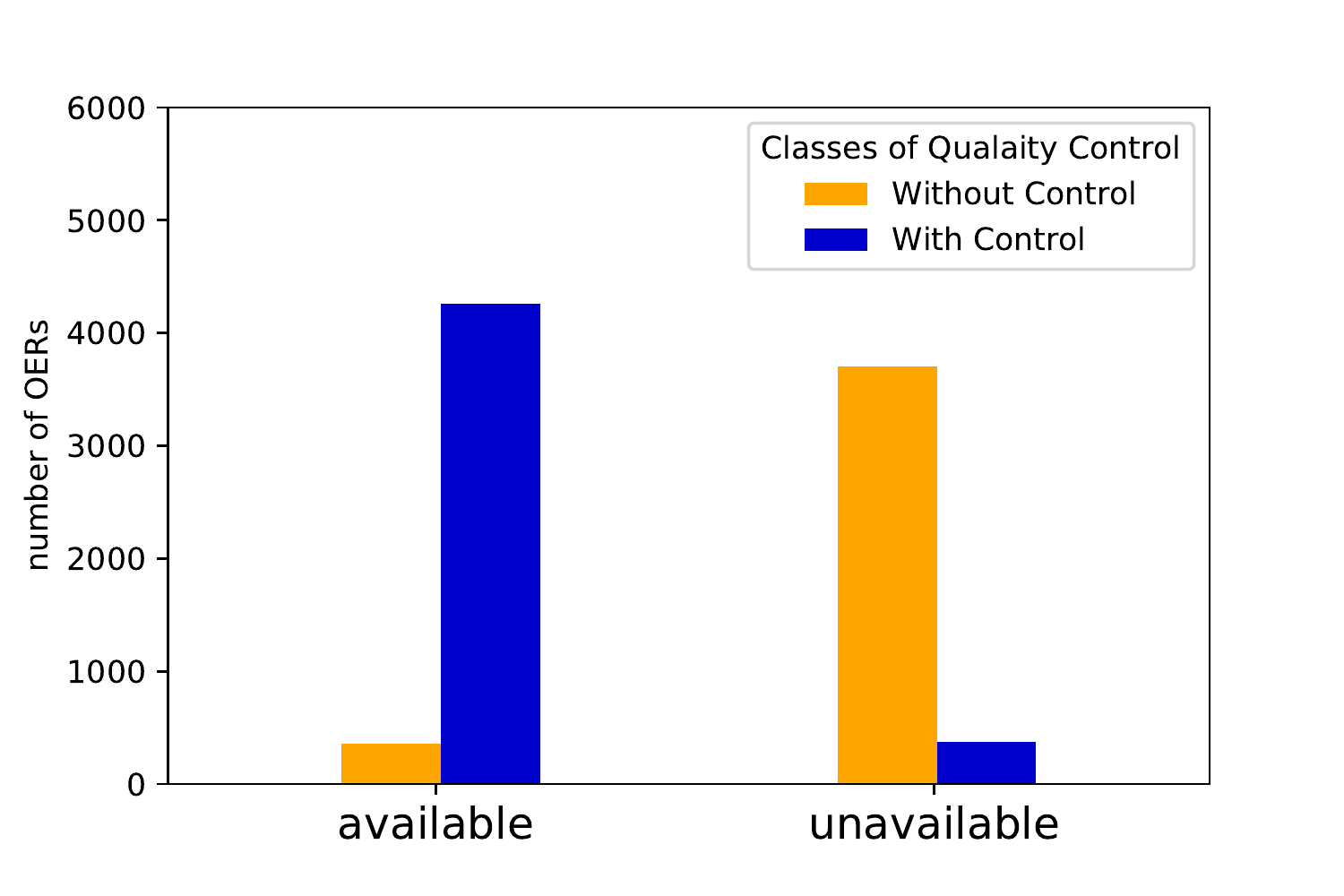}
    \caption{\textit{Language} metadata}
    \label{fig-language}
  \end{subfigure}
  \begin{subfigure}{0.4\textwidth}
    \includegraphics[width=6cm]{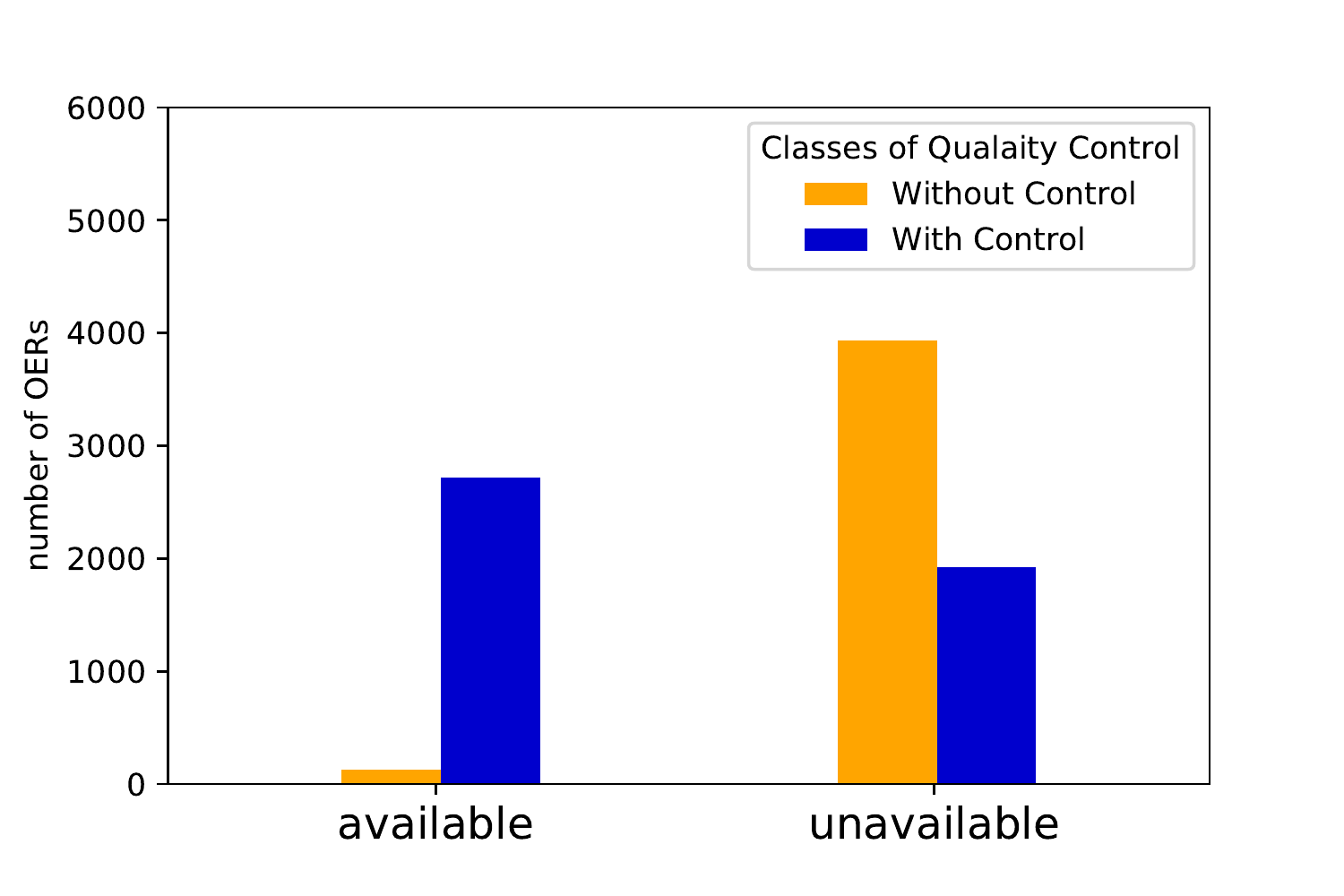}
    \caption{\textit{Time Required} metadata}
    \label{fig-time}
  \end{subfigure}
  \begin{subfigure}{0.4\textwidth}
    \includegraphics[width=6cm]{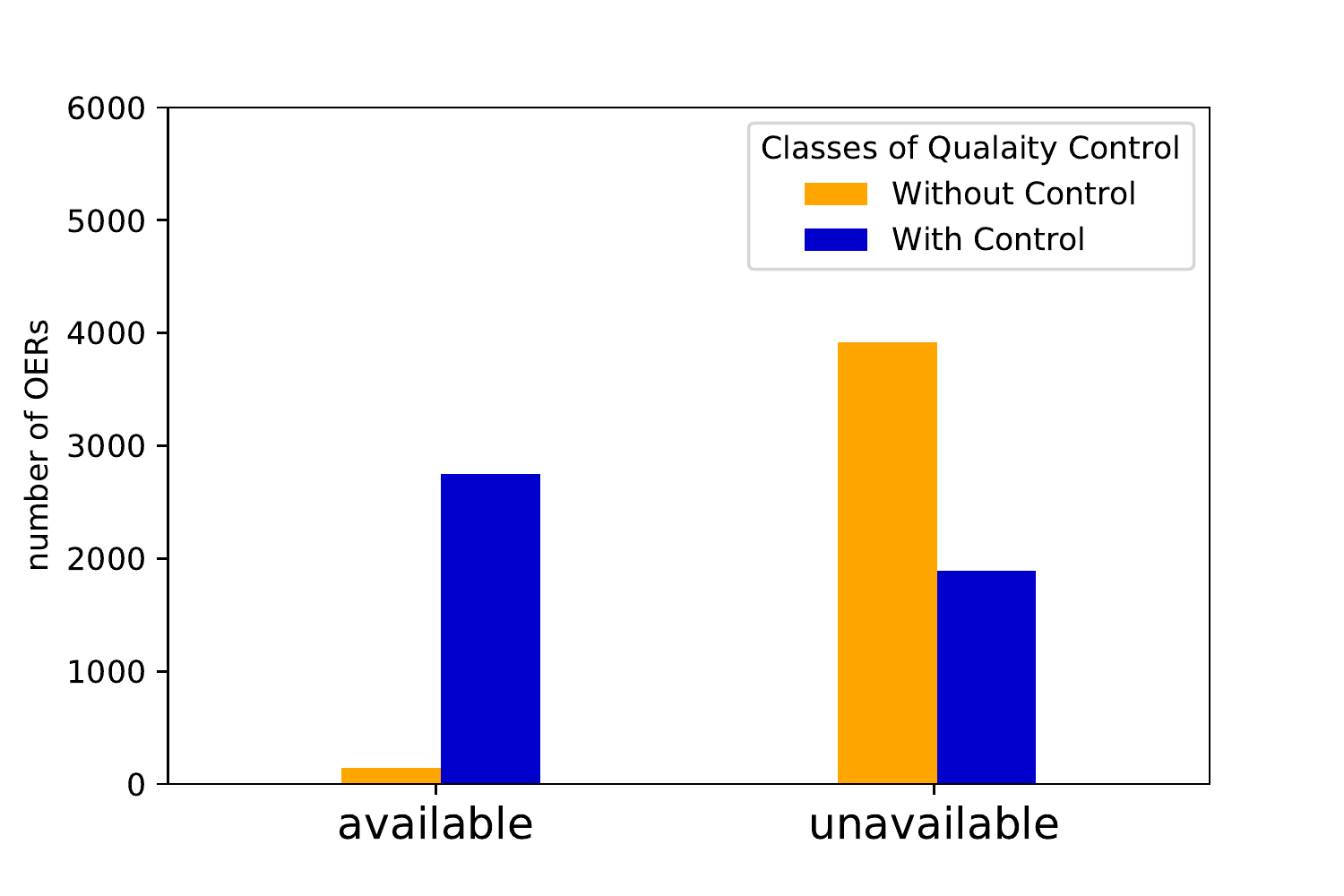}
    \caption{\textit{Accessibilities} metadata}
    \label{fig-acc}
  \end{subfigure}
  \caption{Analyzing metadata availability with respect to manual quality control}
  \label{fig-manual}
\end{figure*}

As a point of departure, we used our Skillscommons dataset to explore the availability of different OER metadata elements (i.e., level, language, time required, accessibilities) based on their quality control categories ("with control" or "without control").
The results of the analysis are summarized in Figure~\ref{fig-manual}:
\begin{itemize}
    \item \textit{Level} refers to the learners’ expertise or educational level in relation to a specific OER. Figure~\ref{fig-level} illustrates how the quality control increases the availability of level metadata.
    \item \textit{Language} refers to the available language versions of an OER. Figure~\ref{fig-language} illustrates the effect of quality control in increasing the availability of language metadata.
    \item \textit{Time Required} refers to the expected duration needed to complete an OER. Figure~\ref{fig-time} shows that it is more likely that OERs with quality control have this type of metadata.
    \item \textit{Accessibilities} defines the accessibility guidelines supported by an OER. Figure~\ref{fig-acc} illustrates how quality control increases availability of the accessibility metadata.
\end{itemize}

\begin{figure}[h]
  \centering
  \includegraphics[width=6.5cm]{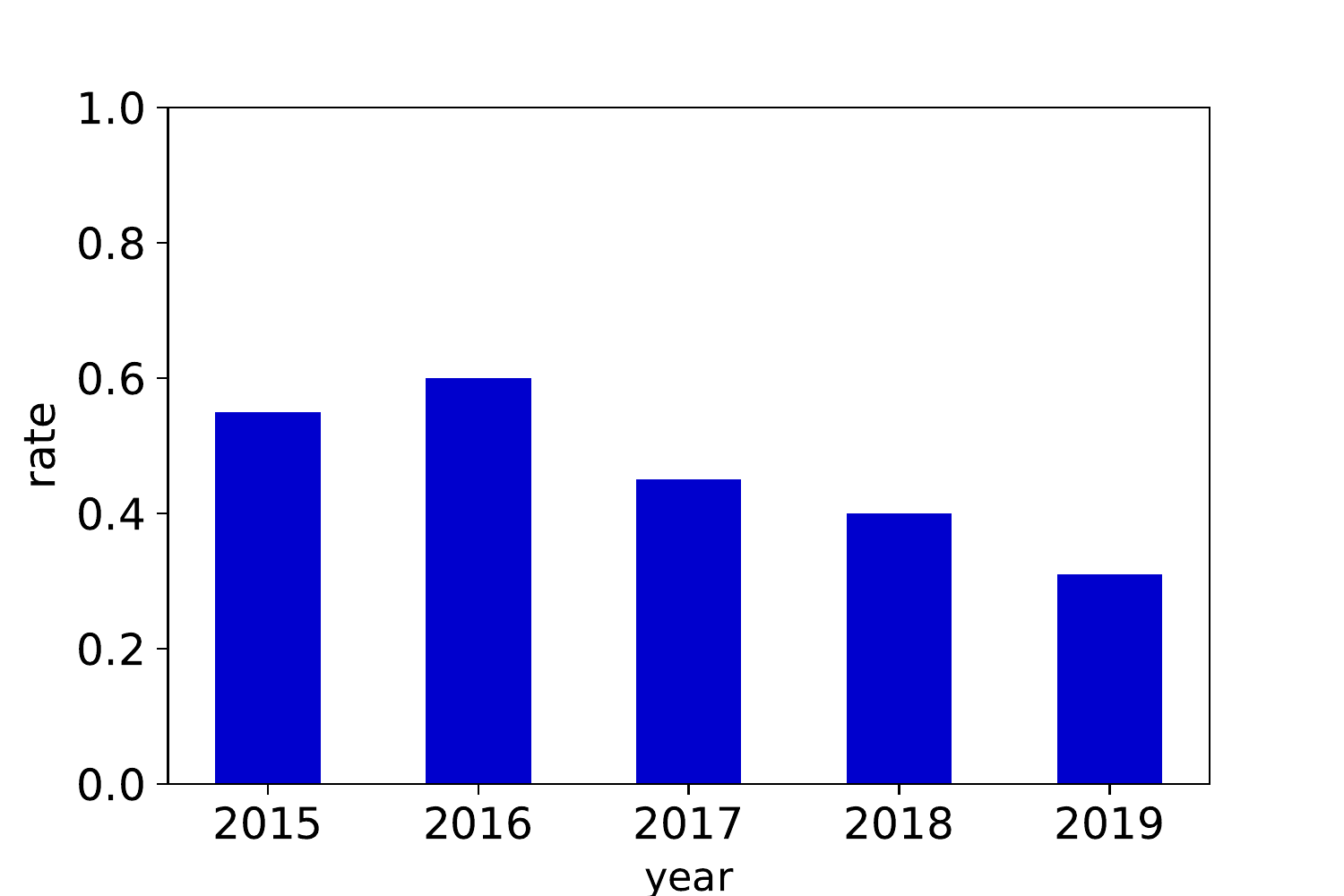}
  \caption{Proportion of manual OER quality control}
  \label{fig-rates}
\end{figure}

To clarify, in each chart, bars on the left show the number of OERs including the particular metadata field, and bars on the right show the number of OERs missing that particular metadata field. Moreover, blue bars are related to the number of OERs with quality control, and orange bars show the number of OERs without quality control. For example, in the left chart of \textit{Level} metadata, you can see more than 4,000 OERs have passed through quality control and also contain Level metadata. At the same time, around 3,000 OERs did not go through quality control, and also do not contain the Level metadata.
The plots in Figure ~\ref{fig-manual} show a clear increase in OER metadata quality (i.e., availability) in the quality controlled OERs, which can be interpreted as a result of OER quality control. However, Figure ~\ref{fig-rates} shows that the proportion of manual OER quality control in our dataset has been decreasing over the last years. We believe that the growing number of OER providers and contents are among the main reasons for this negative change in the proportion of manual OER quality control.

\begin{figure*}
 \centering
 \begin{subfigure}[b]{0.32\textwidth}
    \includegraphics[width=4.7cm]{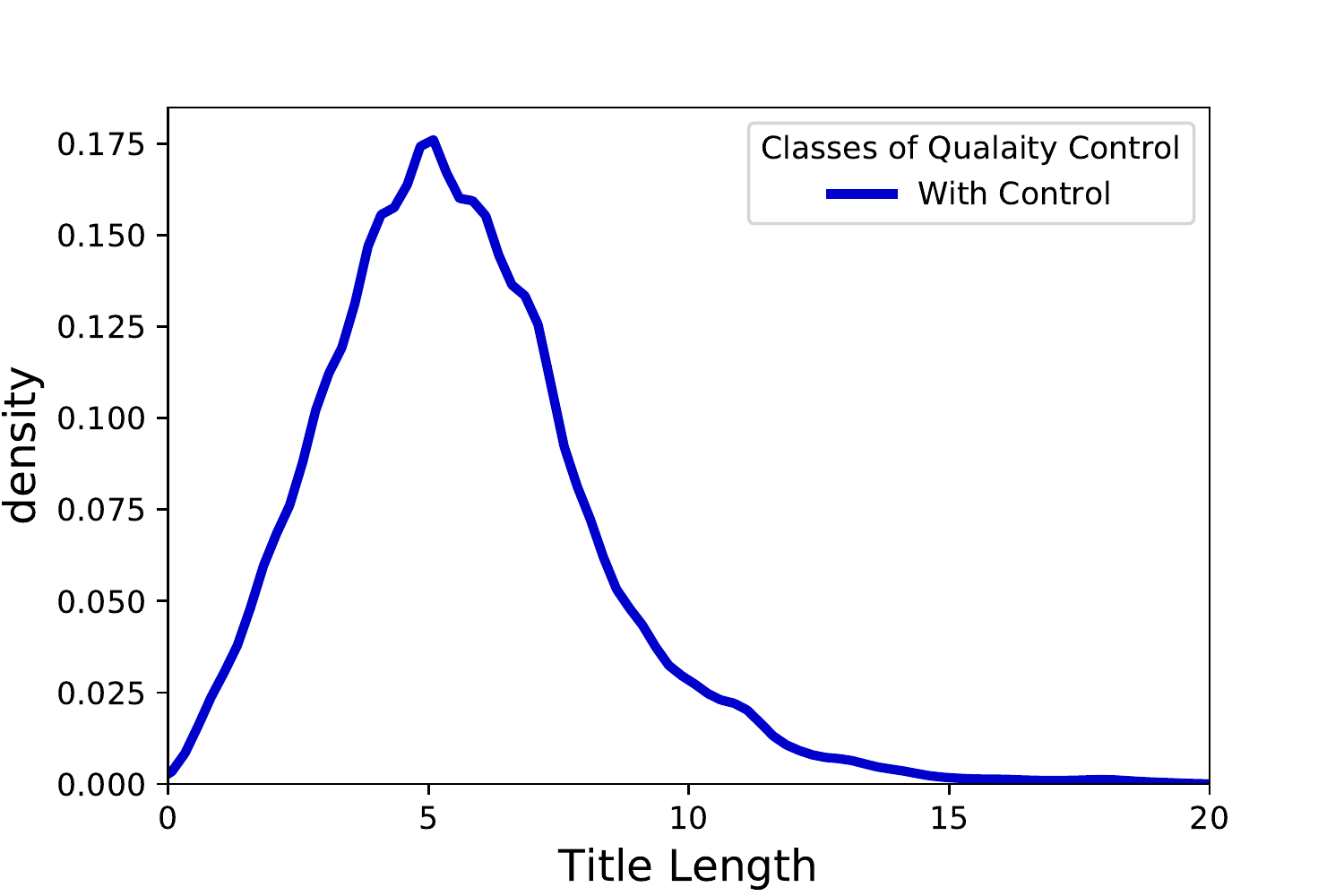}
    \caption{Length of \textit{Title}\\(word count)}
    \label{fig-title}
 \end{subfigure}
 \begin{subfigure}[b]{0.32\textwidth}
    \includegraphics[width=4.7cm]{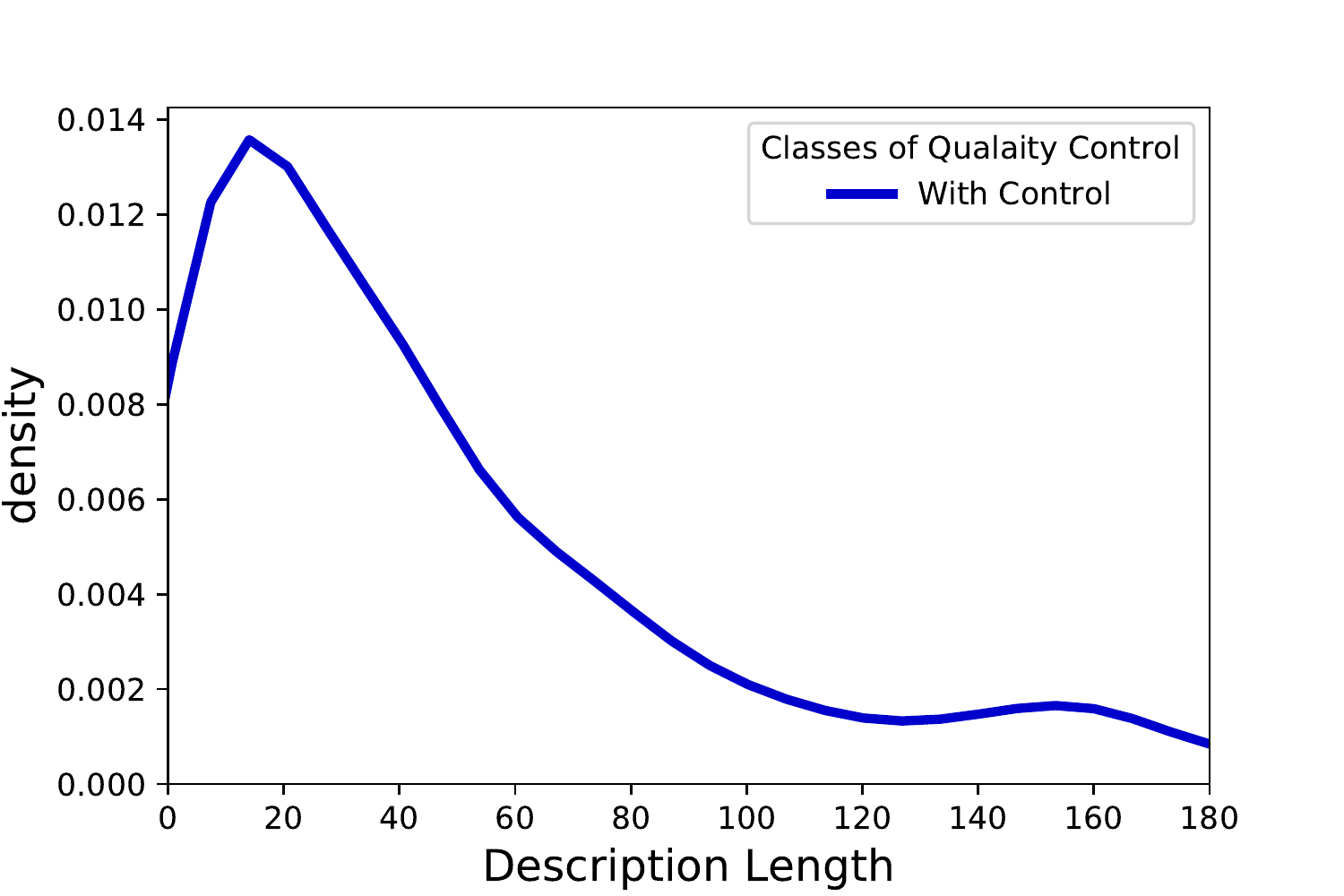}
    \captionsetup{justification=centering}
    \caption{Length of \textit{Description}\\(word count)}
    \label{fig-desc}
 \end{subfigure}
 \begin{subfigure}[b]{0.32\textwidth}
    \includegraphics[width=4.7cm]{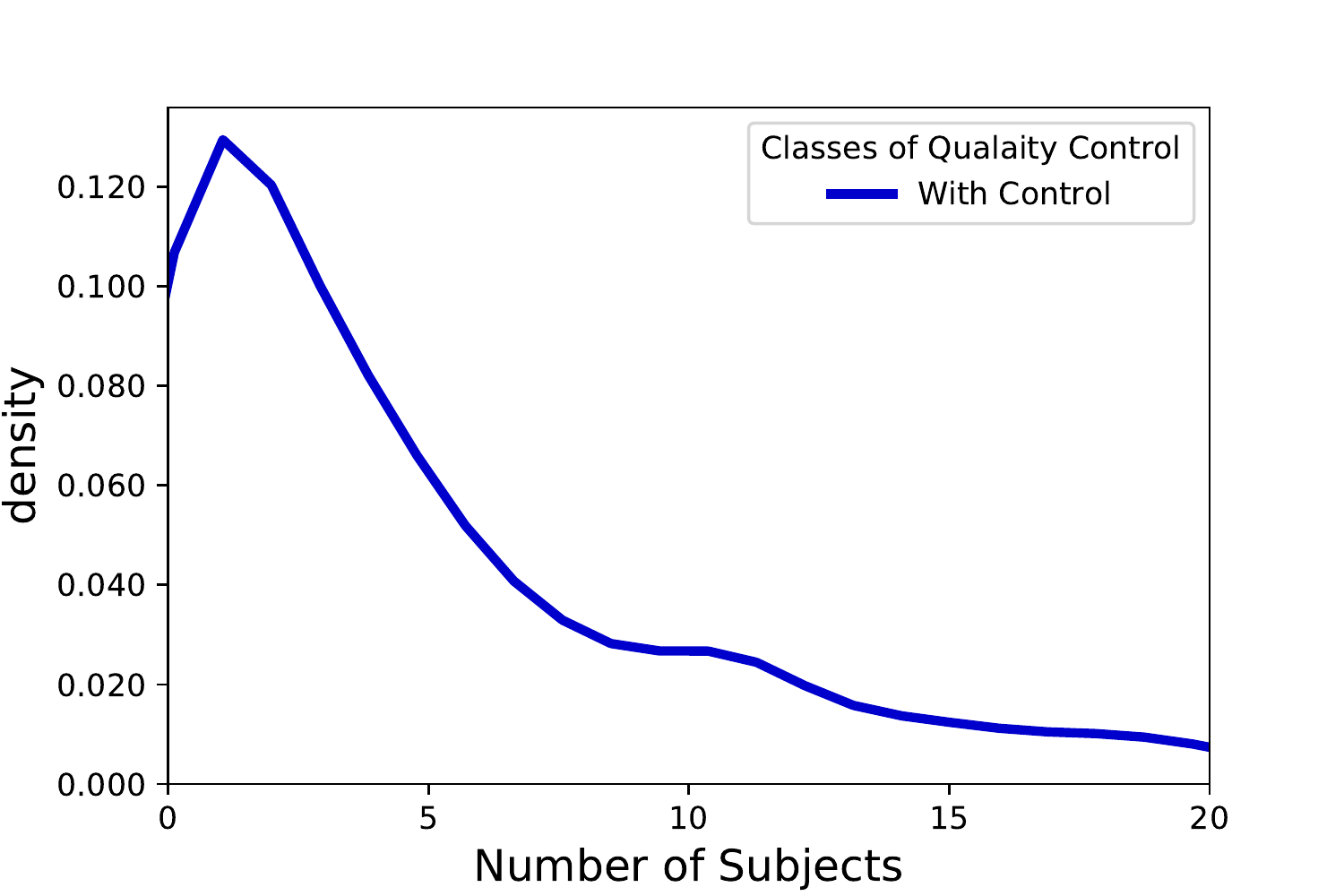}
    \captionsetup{justification=centering}
    \caption{Length of \textit{Subject}\\(subject count)}
    \label{fig-subj}
 \end{subfigure}
 \caption{Metadata analysis of quality controlled OER elements}
 \label{fig-density}
\end{figure*}

As results of our exploratory data analysis, (1) we can use the OER metadata subset with already existing quality control to define quality benchmarks for metadata elements, and (2) it is necessary to define a method that facilitates the automatic assessment of OER metadata quality, and consequently the quality control of OERs.
Therefore, as the next step in our analysis, we focused on OERs with quality control and screened the remaining metadata elements (i.e., title, description, and subjects) of these OERs:
\begin{itemize}
    \item Title refers to the title given to an OER. Figure~\ref{fig-title} shows the distribution of title length (as number of words).
    \item Description refers to the content summary of an OER. Figure~\ref{fig-desc} illustrates the distribution of description length (as number of words).
    \item Subject refers to the subjects (topics) which an OER addresses. Figure~\ref{fig-subj} shows the distribution of subjects (as number of subjects). 
\end{itemize}

The plots in Figure~\ref{fig-density} show that these features have distributions similar to normal. Therefore, it is possible to fit a normal distribution on them and build a scoring model based on the distribution parameters.

\subsection{OER Metadata Scoring Model}
As the first step to build our scoring model, we defined the importance of each metadata field and a rating function based on those OERs, which went through quality control.
For this purpose, we set the importance rate of each metadata field according to its availability rate (between $0$ and $1$) among quality controlled OERs. For instance, all quality controlled OERs have a title and therefore, we set the importance rate of title to $1$, and for language, we set it to $0.92$ since \textit{92\%} of the controlled OERs have language metadata. Accordingly, we normalised the calculated importance rates as \textit{normalized importance rate}.
Afterwards, for each field, we created a rating function based on the OERs with quality control, in order to rate metadata values. The rating function of the fields (title, description, subjects) was devised by fitting a normal distribution on their value length as they have distributions similar to normal, as illustrated in Figure~\ref{fig-density}. We used the reverse of Z-score concept (as $\frac{1}{\lceil |x-\bar{x}|/s\rceil}$ where $\bar{x}$ and $s$ is the mean and standard deviation respectively of the field in the dataset) to rate the metadata values based on the properties of the controlled OERs. Thus, the closer an OER \textit{title/description/subject} length to the mean of distributions, the higher is the rate\footnote{It should be mentioned that when a field value is \textit{equal to the mean} or \textit{empty}, the rate will be $1$ or $0$, respectively.}. Moreover, we used a Boolean function for the four fields (level, length, language, accessibility) which assigns $1$ when they have a value and assigns $0$ otherwise. Table~\ref{tab:tbl-metadata} shows the metadata fields, \textit{importance rates}, \textit{normalized importance rates}, and the \textit{rating functions}.

\begin{table}[h]
\caption{OER metadata fields and importance}\label{tab:tbl-metadata}
\begin{tabular}{|p{2.2cm}|c|c|c|}
\hline
\rowcolor{Gray}\textbf{Type} & $\begin{matrix} \text{\textbf{Importance Rate}} \\ \text{[0-1]} \end{matrix}$ & $\begin{matrix} \text{\textbf{Normalized Importance Rate}} \\ \text{[0-1]} \end{matrix}$ & $\begin{matrix} \text{\textbf{Rating Function}} \\ \text{[0-1]} \end{matrix}$ \\ \hline
Title & 1 & 0.17 & $\frac{1}{\lceil |x-5.5|/2.5\rceil}$ \\  \hline
Description & 1 & 0.17 & $\frac{1}{\lceil |x-54.5|/40\rceil}$ \\  \hline
Subjects & 0.86 & 0.145 & $\frac{1}{\lceil |x-4.5|/3.5\rceil}$ \\  \hline
Level & 0.98 & 0.165 & If available: 1; else: 0 \\  \hline
Language & 0.92 & 0.155 & If available: 1; else: 0 \\  \hline
Time Required & 0.58 & 0.098 & If available: 1; else: 0 \\  \hline
Accessibilities & 0.59 & 0.099 & If available: 1; else: 0 \\  \hline
\end{tabular}
\end{table}

Finally, the following two scoring models were defined to cover the availability and adherence of the defined benchmarks:
\subsubsection{Availability Model} We calculate the availability score of an OER  as the following equation where $norm\_import\_rate(k)$ is \textit{Normalized Importance Rate} of metadata field $k$. This score shows how complete that metadata is in a weighted summation, in which the normalized important rates are the weights. Therefore, the more an OER contains important fields, the higher the availability score is. For instance, an OER which has \textit{title, description}, and \textit{level} (i.e., metadata with high importance rates), achieves a higher availability score than another one which has metadata for \textit{subjects, language, time required}, and \textit{accessibilities}.
\begin{equation} \label{eq:availability}
  avail\_score(o) = \sum_{k=available fields} norm\_import\_rate(k)
\end{equation}
\subsubsection{Normal Model} We calculate the normal score of an OER \textit{o} as the following equation, where $norm\_import\_rate(k)$ is the \textit{Normalized Importance Rate} of metadata field $k$, and \textit{rating(o,k)} is the assigned rating to OER $o$ based on the rating function of metadata field $k$. This score shows how close metadata to the defined benchmark is (based on OERs metadata with quality control). With this scoring model, an OER which has the most similar metadata properties with the metadata of quality controlled OERs, achieves the highest normal score.

\begin{equation} \label{eq:adherence}
  norm\_score(o) = \sum_{k=fields} norm\_import\_rate(k) * rating(o,k)
\end{equation}

\subsection{Predicting the quality of OERs based on their metadata}
We trained a machine learning model to predict the quality of OERs based on their metadata and our scoring model. Therefore, we got the OERs \textit{with control} as higher quality class (containing \textit{4,651} OERs) and set the remaining as lower quality class (containing \textit{4,236} OERs). As a classifier, a Random Forest model was trained on the \textit{SkillsCommons} dataset to build a model that makes a binary decision: high-quality/low-quality.
We used \textit{80\%} of the data as a training set and the remaining \textit{20\%} as test set. The classifier achieved an accuracy of \textit{94.6\%}, where \textit{95\%} of F1-score for \textbf{with control} class, and \textit{94\%} of F1-score for \textbf{without control} class\footnote{The implementation steps and results in Python are available on:~\url{https://github.com/rezatavakoli/ICALT2020_metadata}}. Moreover, we extracted the importance value of each feature for the classification task. Table ~\ref{tbl-features} represents the features of our model and their importance score (i.e. effect) in our model.

\begin{table}[]
\centering
\caption{OER quality prediction model features}
\label{tbl-features}
\begin{tabular}{|p{4.5cm}|c|}
\hline
\rowcolor{Gray}\textbf{Feature} & \textbf{Importance score [0-1]} \\ \hline
Availability Score & 0.32 \\ \hline
Normal Score & 0.25 \\ \hline
Level Metadata Availability & 0.23 \\ \hline
Description Length & 0.10 \\ \hline
Title Length & 0.05 \\ \hline
Subjects Length & 0.05 \\ \hline
\end{tabular}
\end{table}

\section{Validation}\label{sec-validation}
In this section, we report the results of applying our scoring and prediction models on our \textit{Youtube} dataset, including the metadata of \textit{884} educational videos in \textit{32} subjects in the areas of \textit{Information Technology} and \textit{Health Care}.
First, we applied our scoring and prediction models on the dataset to classify the videos into two groups: \textit{with control} (higher quality) and \textit{without control} (lower quality)\footnote{In order to apply our model, we set our required fields based on the video properties. For instance, we set \emph{level availability} based on the videos title, and set \emph{length availability} to "available" as all videos have length metadata.}. After classification, we got \textit{477} videos \textit{with control} and \textit{407} videos \textit{without control}. Then, we needed to identify a metric in their metadata to compare the two groups in order to check whether our model detects the groups of videos with higher quality or not. Therefore, we decided to focus on video \textit{rating} feature as a quality indicator from the users’ perspective, which is calculated based on likes and dislikes, and one of the most commonly used metrics of quality assessment of videos~\citep{moldovan2016vqamap}.
Finally, for each of the \textit{32} subjects, we calculated the average of video ratings for each of the predicted groups (\textit{with control} as higher quality and \textit{without control} as lower quality). Table~\ref{tbl-validation} shows the subjects, the difference of the average rating between the groups, and the difference sign which specifies whether our model predicted correctly and the "with control" group has higher ratings (shows with $+$) or not (shows with $-$).

\begin{table}[b]
\centering
\caption{Difference between videos rating of groups}
\label{tbl-validation}
\begin{tabular}{|p{3cm}|c|c|}
\hline
\rowcolor{Gray}\textbf{Subject} & \textbf{Rating Difference} & \textbf{Difference Sign} \\ \hline
bioethics & 0.15 & + \\ \hline
deep learning & -0.15 & - \\ \hline
infectious disease & 0.14 & + \\ \hline
sleep disorder & -0.14 & - \\ \hline
apache spark & 0.13 & + \\ \hline
data mining & 0.10 & + \\ \hline
allergies & 0.09 & + \\ \hline
vaccinations & 0.08 & + \\ \hline
women and nutrition & -0.08 & - \\ \hline
data management & 0.07 & + \\ \hline
SQL language & -0.06 & - \\ \hline
brain tumors & 0.05 & + \\ \hline
big data & 0.05 & + \\ \hline
cancer prevention & 0.05 & + \\ \hline
data cleaning & 0.05 & + \\ \hline
sun awareness & 0.05 & + \\ \hline
addiction & 0.05 & + \\ \hline
data visualization & 0.04 & + \\ \hline
psychology & 0.03 & + \\ \hline
neural network & 0.03 & + \\ \hline
apache hadoop & 0.03 & + \\ \hline
stress management & 0.02 & + \\ \hline
tensorflow & 0.02 & + \\ \hline
obesity care & 0.02 & + \\ \hline
python language & 0.02 & + \\ \hline
R language & 0.02 & + \\ \hline
statistics & 0.02 & + \\ \hline
text mining & 0.02 & + \\ \hline
machine learning & 0.01 & + \\ \hline
prostate cancer & 0.01 & + \\ \hline
eye care & 0.01 & + \\ \hline
smoking health risks & -0.01 & - \\ \hline
\textbf{Average} & \textbf{0.05} & \textbf{+} \\ \hline
\end{tabular}
\end{table}

As per the results detected by our prediction model, the average rating in a group with higher quality has 0.05 higher video rating than the lower quality group. This is very reasonable considering the standard deviation of ratings in the dataset of \textit{0.25}. To further elaborate, the maximum difference between around \textit{80\%} of the ratings is \textit{0.25}. Therefore, dividing them into two groups with a rating difference of \textit{0.05}, emphasizes that our classifier works well in this context.
Additionally, in \textit{27} out of \textit{32} subjects (\textbf{84.3\%}), where our model detected higher quality groups, they had higher ratings.

\section{Discussion}\label{sec-discussion}
\subsection{OER Metadata}
Based on the exploratory analysis on our OER dataset, it is clear that there is a strong relationship between OER quality control and the metadata quality. Therefore, the more an OER passes the quality control process, the higher the chance of including high-quality metadata is. Accordingly, we can define benchmarks for metadata quality by analyzing the controlled OERs. On the other hand, using metadata quality as a proxy for OER content quality can be beneficial in developing automatic quality control processes for OERs.
According to the analysis of quality controlled OERs, \textit{Title} and \textit{Description} metadata play a key role in publishing OERs, as all of the controlled OERs contain these two fields in their metadata. Moreover, more than \textit{85\%} of the controlled OERs include metadata regarding \textit{Language} and \textit{Level}, and \textit{Subject} which shows the importance of these three fields in defining OERs.

\subsection{Metadata Scoring} Analyzing the importance values in our quality prediction model reveals the effectiveness of our proposed scores for metadata, as the \textit{Random Forest} model assigns the highest value to our \textit{Availability Score} and \textit{Normal Score} features. Therefore, these two proposed indicators illustrate the quality of OER metadata well and can be applied not only for metadata scoring, but also for OER content quality prediction.

\subsection{Quality Prediction Model} The F1-score of our proposed prediction model (\textbf{94.6\%}) shows that we can accurately predict the quality of OERs in \textit{SkillsCommons} repository. Our validation step on \textit{Youtube} dataset also supports the generalizability of our model, which can be applied in different repositories and various types of educational resources (e.g. videos, text-based).
Moreover, according to the result of our validation step, as our prediction model detected the higher quality groups in \textit{14} (out of \textit{16}) \textit{Information Technology} topics and in \textit{13} (out of \textit{16}) \textit{Health Care} topics, the proposed \textit{Random Forest} prediction model works well in different topic areas.

\section{Conclusion and Future Work}\label{sec-conclusion}
In this study, we collected and analyzed the metadata of a large OER dataset to provide deeper insights into OER metadata quality, and proposed a scoring and a prediction model to evaluate the quality of OER metadata and as a consequence OER content quality. We deem that our proposed models not only help OER providers (e.g. repositories and authors) to revisit and think about the importance of the quality of their metadata, but also facilitate the quality control of OERs in general, which is essential in the light of rapidly growing number of OERs and OER providers. Applying our model on the \textit{Skillscommons} dataset indicated that it can detect OERs with quality control with the accuracy of \textbf{94.6\%}. We also validated our approach in another context, by applying our scoring and prediction model to open educational videos on \textit{Youtube}. The results show that our approach successfully detects videos with higher user rating values. The validation step indicates that our approach can be used on different OER repositories.
We consider this study as one of the first important steps to propose intelligent models to improve OER metadata quality and consecutively OER content. In the future, we plan to further improve and validate our models by collecting more data from other repositories and consider more metadata features (e.g. text-based analysis of title and description).

\bibliographystyle{ACM-Reference-Format}
\bibliography{paper}

\end{document}